\documentclass[10pt, conference, compsocconf]{IEEEtran}
\usepackage[pdftex]{graphicx}
\usepackage[cmex10]{amsmath}
\usepackage{array}
\usepackage{color}
\usepackage{listings}
\usepackage{fixltx2e}
\usepackage{url}
\newcommand{\bq}{\begin{equation}}
\newcommand{\eq}{\end{equation}}

\newcommand{\flops}{\mbox{flops}}

\newcommand{\GBS}{\mbox{GB/s}}

\newcommand{\GHZ}{\mbox{GHz}}

\newcommand{\W}{\mbox{W}}

\newcommand{\LUP}{\mbox{LUP}}
\newcommand{\LUPs}{\mbox{LUPs}}

\newcommand{\bytes}{\mbox{bytes}}
\newcommand{\byte}{\mbox{byte}}
\newcommand{\bit}{\mbox{bit}}

\newcommand{\GB}{\mbox{GB}}

\newcommand{\KiB}{\mbox{KiB}}
\newcommand{\MiB}{\mbox{MiB}}

\newcommand{\eos}{~.}

\newcommand{\construction}[1]{}

\usepackage{caption}
\usepackage{float}
\usepackage[lofdepth,lotdepth]{subfig}
\graphicspath{{./figures/}}
\usepackage{sidecap}

\begin{document}
%
\title{Towards  energy efficiency and maximum computational intensity for \\
  stencil algorithms using wavefront diamond temporal blocking}

\author{
\IEEEauthorblockN{Tareq Malas}
\IEEEauthorblockA{
Extreme Computing Research Center (ECRC)\\
King Abdullah University of Science and Technology\\
Thuwal, Saudi Arabia\\
tareq.malas@kaust.edu.sa}\\
\IEEEauthorblockN{Hatem Ltaief}
\IEEEauthorblockA{ 
Extreme Computing Research Center (ECRC)\\
King Abdullah University of Science and Technology\\
Thuwal, Saudi Arabia\\
hatem.ltaief@kaust.edu.sa}

\and
\IEEEauthorblockN{Georg Hager}
\IEEEauthorblockA{
Erlangen Regional Computing Center (RRZE)\\
Friedrich-Alexander University of Erlangen-Nuremberg\\
Erlangen, Germany\\
georg.hager@fau.de}\\
\IEEEauthorblockN{David Keyes}
\IEEEauthorblockA{ 
Extreme Computing Research Center (ECRC)\\
King Abdullah University of Science and Technology\\
Thuwal, Saudi Arabia\\
david.keyes@kaust.edu.sa}
}


%

\maketitle


\begin{abstract}
We study the impact of tunable parameters on computational intensity
(i.e., inverse code balance) and energy consumption of
multicore-optimized wavefront diamond temporal blocking (MWD) applied
to different stencil-based update schemes. MWD combines the concepts
of diamond tiling and multicore-aware wavefront blocking in order to
achieve lower cache size requirements than standard single-core
wavefront temporal blocking. We analyze the impact of the cache block
size on the theoretical and observed code balance, introduce loop
tiling in the leading dimension to widen the range of applicable
diamond sizes, and show performance results on a contemporary Intel
CPU. The impact of code balance on power dissipation on the CPU and in
the DRAM is investigated and shows that DRAM power is a decisive
factor for energy consumption, which is strongly influenced by the
code balance. Furthermore we show that highest performance does not
necessarily lead to lowest energy even if the clock speed is fixed.
\end{abstract}


\section{Introduction}

Regular stencil computations are major contributors to the runtime of many scientific applications. They arise as kernels in structured grid finite-difference and finite-volume discretizations of partial differential equation conservation laws and constitute the principal innermost kernel in many temporally explicit schemes for such problems.  They also arise as a co-principal innermost kernel of Krylov solvers for temporally implicit schemes on regular grids. In iterative stencil computations, each point in a multi-dimensional
spatial grid is updated using weighted contributions from its
neighbor points, defined by the stencil operator.  The stencil
operator specifies the relative coordinates of the contributing points
and their weights.  The weights can be constant or variable in space and/or time with some
or no symmetry to be exploited around the updated point.  The grid update operation
over the complete spatial domain (one ``sweep'') is repeated over many time steps (or iterations).

A  high bytes-per-flop requirement is a prominent property of many stencil computations. It can be quantified by means of the \emph{code balance}
metric,
i.e., the number of bytes transferred over a relevant bottleneck (usually
the memory interface) divided by the ``work'' that can be done using this
data. Since the number of \flops\ may vary in stencil computations
due to different formulations of the loop kernel or even compiler
optimizations, the ``lattice site update'' (\LUP) is a better
metric for quantifying work. Hence, the code balance is measured
in $\bytes/\LUP$. Its inverse is commonly called ``computational intensity.''
These metrics can be used to predict the performance of a loop code
in a bandwidth-bound scenario~\cite{Williams09}: the maximum
memory-bound performance is the ratio of maximum memory bandwidth
and the code balance.

At a large code balance, the increasing gap between computation and memory performance in contemporary and future high performance computing systems results in low hardware (i.e., CPU) utilization.  The development of algorithms that can run stencil computations efficiently by reducing the code balance is thus essential for making better use of the hardware. At the same time, energy consumption and power dissipation are getting more attention in scientific computing due to the 
increasing energy and infrastructure cost for 
large systems. Performance and energy considerations are strongly 
intertwined, and any advancement in understanding the former will
also help in controlling the latter.

\subsection{Tested stencil cases}\label{sec:stencils}
We perform our model validation and energy analysis using three ``corner case'' stencils: the 7-point constant-coefficient stencil in Listing~\ref{lst:7const}, which operates at two domain-sized arrays to perform the Jacobi-like update, the 7-point variable-coefficient stencil in Lst.~\ref{lst:7var}, which loads and caches an additional 7 domain-sized coefficient arrays, and the 25-point variable-coefficient stencil in Lst.~\ref{lst:25var}, which operates on 13 coefficient arrays. 
These stencils are corner cases in the sense of including short- and long-range stencils and constant- and variable-coefficient stencils.
Compared to the short-range stencils, the stencil operator of the long-range stencils includes more grid points (larger ``stencil radius'') and has data dependency over more distant points in space from the updated lattice site, which adds more challenges for temporal blocking techniques.
The variable-coefficient stencils have a several times higher data requirement per grid point compared to the constant-coefficient stencils, as they have to load the coefficient arrays, causing more cache pressure when using blocking techniques.

\begin{lstlisting}[float=tb,label=lst:7const,caption={$1^{st}$-order-in-time 7-point constant-coefficient isotropic stencil in three dimensions, with symmetry.}]
#pragma omp parallel for
for(int k=1; k < N-1; k++) {
 for(int j=1; j < N-1; j++) {
  for(int i=1; i < N-1; i++) {
   U[k][j][i] = c0 * V[k][j][i]
     + c1 * ( V[ k ][ j ][i+1]+ V[ k ][ j ][i-1])
     + c1 * ( V[ k ][j+1][ i ]+ V[ k ][j-1][ i ])
     + c1 * ( V[k+1][ j ][ i ]+ V[k-1][ j ][ i ]);
}}}
\end{lstlisting}

\begin{lstlisting}[float=tb,label=lst:7var,caption={$1^{st}$-order-in-time 7-point variable-coefficient stencil in three dimensions, with no coefficient symmetry.}]
#pragma omp parallel for
for(int k=1; k < N-1; k++) {
 for(int j=1; j < N-1; j++) {
  for(int i=1; i < N-1; i++) {
   U[k][j][i] = C0[k][j][i] * V[k][j][i]
     + C1[k][j][i] * V[ k ][ j ][i+1]
     + C2[k][j][i] * V[ k ][ j ][i-1]
     + C3[k][j][i] * V[ k ][j+1][ i ]
     + C4[k][j][i] * V[ k ][j-1][ i ]
     + C5[k][j][i] * V[k+1][ j ][ i ]
     + C6[k][j][i] * V[k-1][ j ][ i ];
}}}
\end{lstlisting}

\begin{lstlisting}[float=tb,label=lst:25var,caption={$1^{st}$-order-in-time 25-point variable-coefficient anisotropic stencil in three dimensions, with symmetry across each axis.}]
#pragma omp parallel for
for(int k=4; k < N-4; k++) {
for(int j=4; j < N-4; j++) {
for(int i=4; i < N-4; i++) {
 U[k][j][i] = C00[k][j][i] * V[k][j][i]
  +C01[k][j][i]*(V[ k ][ j ][i+1]+V[ k ][ j ][i-1])
  +C02[k][j][i]*(V[ k ][j+1][ i ]+ V[ k ][j-1][ i ])
  +C03[k][j][i]*(V[k+1][ j ][ i ]+ V[k-1][ j ][ i ])
  +C04[k][j][i]*(V[ k ][ j ][i+2]+ V[ k ][ j ][i-2])
  +C05[k][j][i]*(V[ k ][j+2][ i ]+ V[ k ][j-2][ i ])
  +C06[k][j][i]*(V[k+2][ j ][ i ]+ V[k-2][ j ][ i ])
  +C07[k][j][i]*(V[ k ][ j ][i+3]+ V[ k ][ j ][i-3])
  +C08[k][j][i]*(V[ k ][j+3][ i ]+ V[ k ][j-3][ i ])
  +C09[k][j][i]*(V[k+3][ j ][ i ]+ V[k-3][ j ][ i ])
  +C10[k][j][i]*(V[ k ][ j ][i+4]+ V[ k ][ j ][i-4])
  +C11[k][j][i]*(V[ k ][j+4][ i ]+ V[ k ][j-4][ i ])
  +C12[k][j][i]*(V[k+4][ j ][ i ]+ V[k-4][ j ][ i ]);
}}}
\end{lstlisting}

\subsection{Contribution}

This work  makes the following relevant contributions:
\begin{itemize}
\item We introduce a traffic model for stencil codes optimized with
  multicore wavefront diamond temporal blocking.  The model predicts
  the data volume over the memory bus, and thus the computational
  intensity, on a multicore processor.
\item We show by direct measurements that the model is correct as long
  as the required cache block size is within half the available cache
  size. This means that the MWD technique is able to attain the 
  predicted memory traffic reductions.
\item We show by direct measurements on the Intel Ivy Bridge processor
  that energy to solution for the considered optimized stencil codes
  correlates strongly with execution time, but that this apparently
  simple dependency is the result of two counteracting effects:
  a weak dependence of CPU power on
  the code performance and  a strong dependence of DRAM power on the
  memory traffic.
\end{itemize}

\section{Background}

\subsection{Multicore wavefront diamond temporal blocking}

We perform energy and code balance analysis in this paper using our proposed approach in \cite{malas_mws_sisc2014}. It combines the concepts of diamond tiling and multi-core aware wavefront temporal blocking to construct Multi-core Wavefront Diamond blocking (MWD) for optimizing practically relevant stencil algorithms. 

Our approach builds on a technique introduced by Strzodka \textit{et al.}\ ~\cite{Strzodka:2011:CAT}. They combine diamond tiling with single-threaded wavefront temporal blocking. Our approach replaces the single-core wavefront with the multi-core wavefront proposed by Wellein \textit{et al.}~\cite{wellein5254211}, which provides additional dimension of concurrency and offers large reduction in the cache block size and memory bandwidth requirements, as we shown in the results of our previous work \cite{malas_mws_sisc2014}. 

\begin{figure}[btp]
    \centering
    \includegraphics[width=\columnwidth]{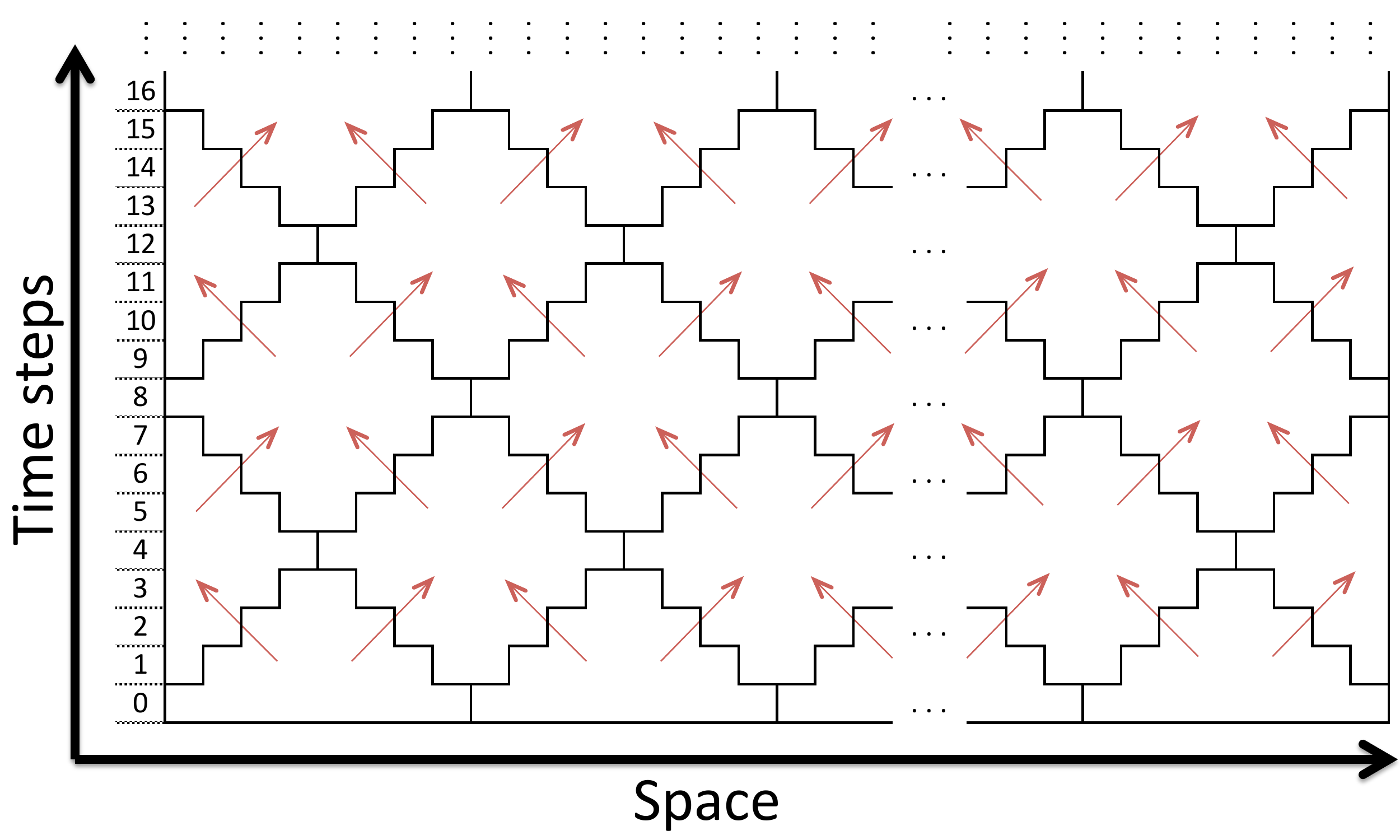}
    \caption{Diamond tiling on a one-dimensional space grid, with arrows
  representing inter-tile data dependencies.  The number of diamond
  tiles per row represents the maximum attainable concurrency,
  as the tiles in the row can be executed independently of each other.}
    \label{fig:diamond_tiling_1d}
\end{figure}

In Figure~\ref{fig:diamond_tiling_1d} we illustrate the concept of diamond tiling for a one-dimensional 3-point stencil. Arrows represent the data dependency across the diamond tiles.
Diamond tiling provides convenient and unified data structure to maximize the in-cache data reuse~\cite{orozco2009ICPP}, has low synchronization requirements, allows concurrent diamond tiles update, and can be utilized to perform domain decomposition in distributed memory setup. 

\begin{figure}[tbp]
    \centering
    \includegraphics[width=\columnwidth]{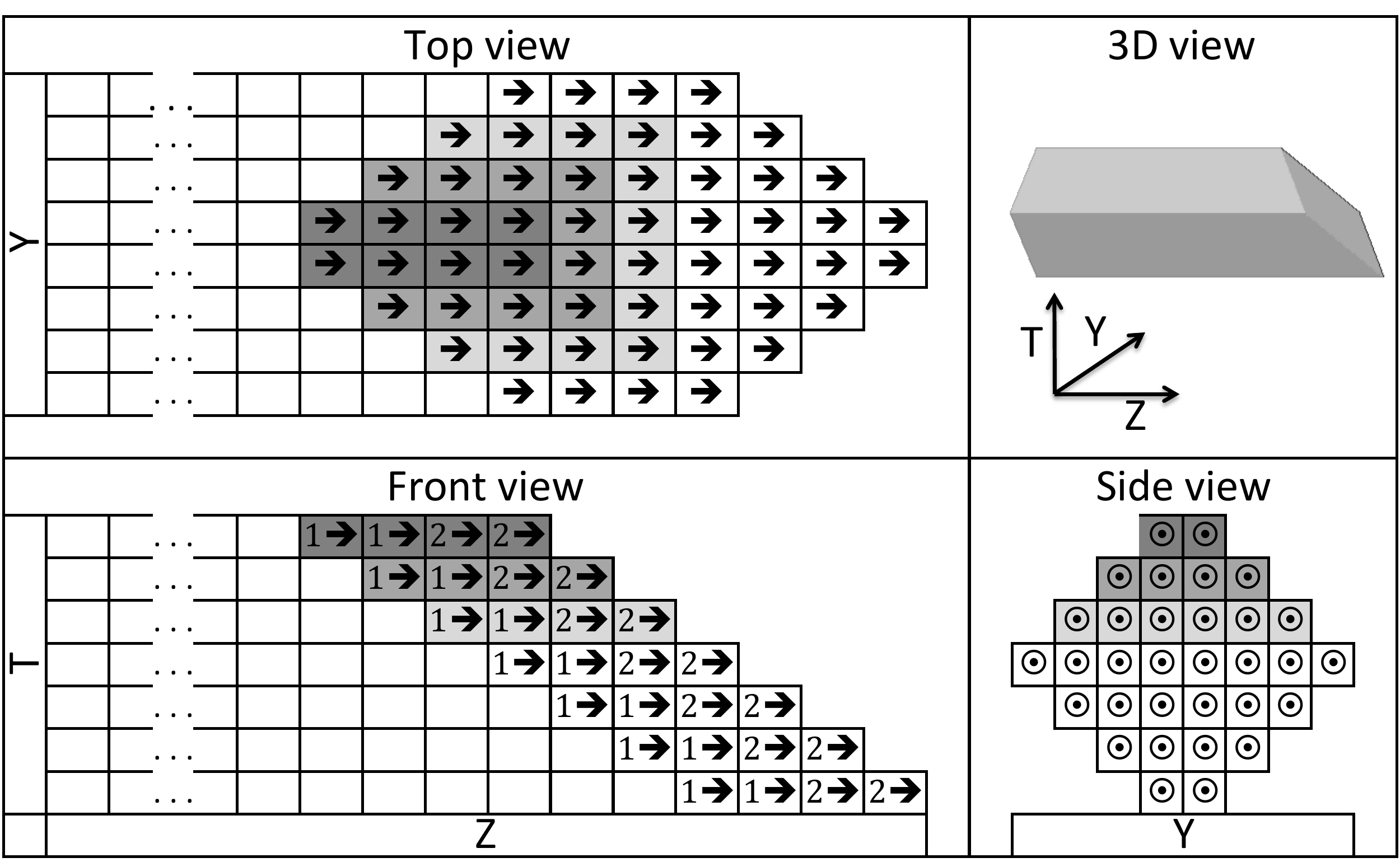}
    \caption{Diamond tiling with a multi-thread wavefront (shown here with 
    two threads) in a three-dimensional space grid, using two frontlines    
    per thread.}
    \label{fig:mwd_2d}
\end{figure}

The MWD space-time tile has the shape of an extruded diamond, as shown in Figure~\ref{fig:mwd_2d}.  The ``frontlines'' parameter determines the number of updated grid points in the wavefront direction per thread per time step in the wavefront update. The fading gray
color represents recently updated grid points, with the darkest
assigned to the most recent update. The wavefront traversal is performed along the $z$ dimension (outer dimension) and the diamond tiling is performed across the $y$ dimension (middle dimension). The $x$ dimension (which is represented by single point in the figure) is left intact to have more contiguous memory accesses for efficient hardware data prefetching and reduced TLB misses.

Threads are assigned to the extruded diamonds in groups (``thread 
groups''), similar to \cite{wellein5254211}. Multiple thread 
groups can run concurrently, updating different diamond tiles and
observing inter-diamond dependencies. The thread group size parameter provides controllable 
tradeoff between concurrency and sharing of the loaded data from memory 
among the threads of the multi-processor. Diamond tiles are dynamically scheduled to the available threads.  A FIFO queue keeps track of the available
diamond tiles for updating.  Threads pop tiles from this queue
to update them.  When a thread completes a tile update, it pushes 
to the queue its dependent diamond tile, if that
has no other dependencies.  The queue
update is performed in an OpenMP critical region to avoid race
conditions.  Since the queue updates are performed infrequently, the
lock overhead is negligible.

Selecting the diamond tile size and the number of frontlines updates is achieved through auto-tuning to achieve the best performance.  To shorten the auto-tuning process, the parameter search space is
narrowed down to diamond tiles that fit within a predefined cache size
range. Several constraints are considered in selecting the auto-tuning test points, for example, having sufficient concurrency and integer number of diamond tiles in each row of diamond tiles.

\subsection{Performance and energy consumption on multicore processors}

Power dissipation, energy to solution, and more advanced
energy-related metrics have become additional optimization targets in
high performance computing besides pure time to solution and resource
efficiency. Fortunately, low time to solution often 
leads to low energy to solution when looking at a fixed set of
resources such as a multicore CPU or a compute node.  Modern
multi-core processors have advanced power gating mechanisms that make
their power dissipation highly dynamic and code-dependent.  For
instance, idle cores (or parts of them) can be put into deep sleep
states. Nevertheless there are successful attempts to describe their
power behavior using simple models, which are not perfectly accurate
but provide useful
insights~\cite{Choi:2013:RME:2510661.2511392,CPE:CPE3180}. We briefly
summarize here the consequences of the energy model derived in
\cite{CPE:CPE3180}.  It assumes a non-zero ``baseline'' or ``static
power,'' which is the extrapolated power dissipation with all cores
idling, and a constant per-core, code-dependent dynamic power
contribution from every core that is not idle:
\bq
W = W_\mathrm{stat} + n\cdot W_\mathrm{dyn}\eos
\eq
One crucial property that influences energy consumption is the scaling
behavior of a code (or rather an execution phase, which is often a
loop nest) across the cores of a multicore chip. If performance scales
across the cores, energy to solution is minimal when using all cores.
On the other hand, if the code performance saturates, like with
strongly memory-bound loop kernels or imbalanced workload, lowest
energy is achieved when using as many cores as required to saturate,
but not more. If the saturation is caused by a hardware
bottleneck (typically memory bandwidth), reducing the clock speed
at larger core count will further decrease the energy to solution.

Given the plethora of relevant stencil variants and possible
optimization techniques, it is impossible to pinpoint a generic energy
behavior for ``stencils at large,'' since those cover the complete
spectrum from fully scalable to strongly saturating. Modern hardware
structures beyond the CPU chip with dynamic power behavior such as
large DRAMs complicate matters further. In this work we show, using
baseline and optimized implementations of three stencil-based
algorithms, that the usual law of ``faster code uses less energy'', or ``race-to-halt'' as defined by Hennessy and Patterson~\cite{hennessy2012computer}, is often but not always true.

\section{Analysis and optimization of the MWD algorithm}

\subsection{Loop tiling for the leading dimension}

The experiments of our MWD implementation in~\cite{malas_mws_sisc2014} showed a degradation in the thread scaling performance of the 7-point variable-coefficient stencil at grid size $N\!=\!680^3$, where the performance at 10 threads was lower than at 8 and 9 threads. Using half the leading dimension ($N_x$) in other experiments (i.e., $340\times\! 680 \times\! 680$) led to good performance scaling up to 10 threads.
This is a result of reducing the cache block size to half, since $N_x$ contributes linearly in the cache block size, as will be shown in Section~\ref{sec:mem_model}.
Blocking in the leading dimension was thus considered to resolve the cache capacity issue. Reducing the leading dimension size can allow for more in-cache data reuse by fitting larger diamond tiles in the cache. This is especially useful when the memory bandwidth remains a bottleneck even with MWD.

To avoid adding more complexity in scheduling tiles to thread groups, the same thread group updates all the tiles in the leading dimension of its assigned extruded diamond sequentially.
The implementation is kept simple by using parallelepiped tiles in the leading dimension.

The performance-optimal tile size in the leading dimension is problem-dependent, as small blocks can result in increased TLB misses and excess data volume due to  hardware prefetching. To handle this issue, we incorporated the tile size selection in our auto-tuning implementation, to tune it along with the diamond tile size and the number of frontlines. 

This tiling implementation can be found in the current code release of our framework \cite{malas_girih_ipdps_2014}.

\subsection{Cache block size model}\label{sec:blk_size_model}
We consider the MWD wavefront cache block size model from~\cite{malas_mws_sisc2014} to validate its correctness and study its impact on the code balance at different diamond sizes.
The model calculations require four parameters: the diamond width $D_w$ in the $y$ axis, the wavefront frontlines number $N_{F}$, the bytes number in the leading dimension $N_{xb}$, the stencil radius $R$, and the number of domain-sized streams in the stencil operator, $N_D$. 
Examples of stencil radius are $R\! =\! 1$ and $R\! =\!4$ at the 7- and 25-point stencils, respectively.
The 7-point constant-coefficient stencil has $N_D=2$ (Jacobi-like update). The 7-point variable-coefficient stencil uses seven
additional domain-sized streams to hold the coefficients.
For a stencil with $R\! =\! 1$, the wavefront width $W_w$ has the size: $W_w\!=\!D_w\!+\!N_{F}\!-\!2$ and the total required bytes in the wavefront cache block $C_S$, with some approximations, is:
\bq\label{eq:7pt_cache_blk_size}
C_S = N_{xb} \cdot \left[N_D\cdot \left(\frac{D_w^2}{2}\!+\! D_w\! \cdot\!(N_{F}\!-\!1)\right) 
  + 2 \cdot(D_w + W_w)\right] \eos
\eq
Here, $N_{xb}$ is the size
of the leading dimension tile size, and ${D_w^2}/{2} + D_w\cdot (N_{F}-1)$ is the
diamond area in the $y$-$z$ plane as shown in the top
view of Figure~\ref{fig:mwd_2d}.
The halo region of the wavefront (i.e., the read-only grid points around the cache block) is $2 \cdot(D_w + W_w)$.

For example, we have $D_w=8$ and $N_{F}=4$ in Figure~\ref{fig:mwd_2d}, so 
$W_w = 8+4-2 = 10$ and the total block size at 7-point constant-coefficient stencil is $N_{xb} \cdot ( 2\cdot (8^2/2 + 8\cdot 3) + 2 \cdot(8 + 10)) = 148\cdot N_{xb}$~\bytes.

The steeper wavefront in higher-order stencils results in different wavefront lengths ($W_w= D_w - 2\cdot
R + N_{F}$) and different $C_S$ as follows:
\bq\label{eq:cache_blk_size}
C_S\!=\!N_{xb}\cdot\!\left[\!N_D\!\cdot\!D_w\!\cdot\! \left(\frac{D_w}{2}\!-\!R\!+\!N_{F}\right) 
  + 2 R (D_w + W_w)\right]\eos
\eq

It is worth mentioning that each thread group requires a dedicated $C_S$ in the blocked cache level. For example, using 1WD in a 12-core Intel Ivy Bridge socket requires fitting $12\cdot C_S$ \bytes\ in the L3 cache.

\subsection{Memory traffic model}\label{sec:mem_model}


In order to validate the effectiveness of the
bandwidth pressure reduction on the memory interface, 
we set up a model to estimate the code balance for the temporally
blocked case. If the wavefront fits completely in the L3
cache, each grid point is loaded once from main memory and is stored once after updating it during the
extruded diamond update.  In this case, the
amount of data transfers during the extruded diamond update consists of  $(2D_w-2)$ data writes plus $(N_D \cdot D_w+2)$ data reads, all multiplied by $N_z$. The number of total \LUPs\ performed through the diamond volume is:
$N_z \cdot D^2_w/2$. The code balance at double precision of a stencil with $R=1$ is thus:
\bq \label{eq:mwd_bw_req}
B_\mathrm{C} = \frac{16 \cdot \left[(2D_w-2)+
    (N_D \cdot D_w + 2)\right]}{D^2_w}\frac{\bytes}{\LUP}\eos
\eq

When $R\!>\! 1$ the amount of data transfers becomes $N_z \cdot \left[(2D_w-2R) +(N_D \cdot D_w +2R)\right]$ and the extruded diamond volume becomes $N_z \cdot D^2_w/(2\cdot R )$. In total, the equation becomes:

\bq \label{eq:mwd_bw_req_ho}
B_\mathrm{C} = \frac{16R \cdot \left[(2D_w-2R)+
    (N_D \cdot D_w + 2R)\right]}{D^2_w}\frac{\bytes}{\LUP}\eos
\eq

\section{Results}

\subsection{Test system and tools}
All benchmark tests were performed on a cluster of dual-socket Intel
Ivy Bridge (Xeon E5-2660v2) nodes with a nominal clock speed of
2.2\,\GHZ\ and ten cores per chip.  This processor has a
maximum thermal design power (TDP) of 95\,\W.
The ``Turbo Mode'' feature was
disabled. Each CPU has a 25\,\MiB\ L3 cache which is shared among all
cores, and core-private L2 and L1 caches of 256\,\KiB\ and 32\,\KiB,
respectively.  All data paths between the cache levels are
half-duplex, 256-\bit\ wide buses, so the transfer of one
64-\byte\ cache line between adjacent caches takes two CPU cycles. The
core architecture supports all Intel Single Instruction Multiple Data
(SIMD) instruction sets up to AVX (Advanced Vector Extensions).
With AVX, one core is able to sustain one full-width (32\,\byte) 
load and one half-width (16\,\byte) store per cycle. In addition,
one AVX multiply and one AVX add instruction can be executed per cycle.
Since one AVX register can hold either four double precision (DP) or
eight single precision (SP) operands, the peak performance of one
core is eight \flops\ per cycle in DP or sixteen \flops\ per cycle in 
SP. 

Each node is equipped with 64\,\GB\ of DDR3-1600 RAM per socket and
has a maximum attainable memory bandwidth of $b_\mathrm S\approx 40\,\GBS$ per
socket (as measured with the STREAM COPY \cite{stream} \cite{McCalpin1995} benchmark).
The nodes are connected by a full non-blocking, fat-tree QDR
InfiniBand network.

For compiling and linking, the Intel C compiler in version 13.1.3 was
used. 
Hardware performance
counter measurements were done with \texttt{likwid-perfctr}
from the LIKWID multicore tools collection \cite{likwid}.
Apart from standard metrics,
\texttt{likwid-perfctr} can also measure power dissipation
and energy consumption
based on the RAPL (Running Average Power Level) mechanism. 
RAPL is an
energy model implemented in hardware with high degree of
accuracy~\cite{10.1109/MM.2012.12}. Its technology allows to measure energy
seamlessly by using hardware counter technology available
on Intel Sandy/Ivy Bridge lines of multicore processors.
On the system used for the tests, RAPL is able to report
CPU energy separately from DRAM energy.
Note that RAPL has been designed not just for monitoring 
instantaneous energy
consumption but also for capping the total energy and power
at the software level.

\subsection{Code balance}
In this section we verify the correctness of our memory traffic and cache block size models. The three stencils described in Section~\ref{sec:stencils} are used at realistic grid sizes.

We use 10WD in our verification experiments, because
it has the largest range of diamond tile sizes that fits in the L3 cache, where all the threads work in a single cache block.
Smaller thread group sizes result in similar behavior, but run out of cache at smaller diamond sizes, as they require one cache block per thread group to co-exist in the cache memory. Hence, we omit those results since they do not add insight beyond 10WD.
The number of frontlines is fixed to 10 across all experiments, which is the minimum allowed to run a 10-thread wavefront in our implementation.

\begin{figure*}[tbp]
    \newcommand*{\sfwidth}{5.5cm}
    \centering
    \subfloat[7-point constant-coefficient stencil using grid size $N=960^3$]{
        \centering
        \includegraphics[width=\sfwidth]{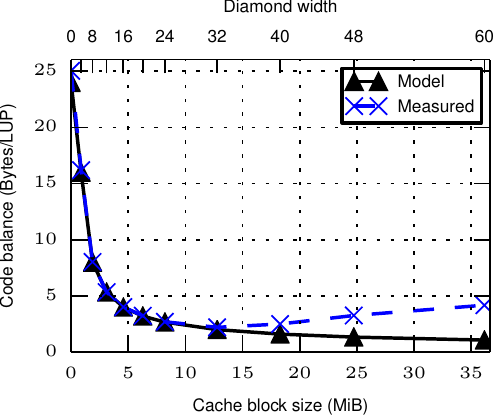}
        \label{fig:7pt_const_code_balance}
	}
	\enskip
    \subfloat[7-point variable-coefficient stencil using grid size $N=680^3$]{
        \centering
        \includegraphics[width=\sfwidth]{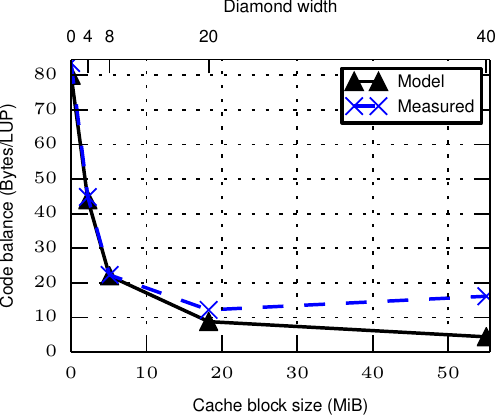}
        \label{fig:7pt_var_code_balance}
	}
	\enskip
	\subfloat[25-point variable-coefficient stencil using grid size $N=480^3$]{
        \centering
        \includegraphics[width=\sfwidth]{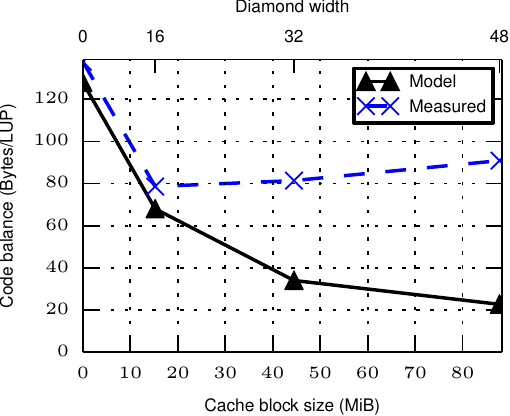}
        \label{fig:25pt_var_code_balance}
	}
	\caption{Cache block size vs. modeled code balance and measured code balance, evaluated at several diamond tile sizes using 10WD with three corner-case stencil operators.}
    \label{fig:code_balance}
\end{figure*}

Figure~\ref{fig:code_balance} shows cache block size vs. code balance at different diamond tile sizes, along with the corresponding diamond width (top $x$ axis).
The coordinates of each data point in the ``Model'' data sets is computed by evaluating the cache size model and code balance model at a given diamond width. The coordinates in the ``Measured'' data sets uses the actual measured code balance in place of the code balance model at the $y$ axis.
Multiples of valid diamond sizes are used in the measurements, where multiples of 4 and 16 are used for the 7-point and 25-point stencils, respectively.
Several diamond sizes are omitted in the figures because they would require a non-integer number of tiles in the diamond-tiling dimension. For example, $D_w=12$ is omitted at Fig.~\ref{fig:7pt_var_code_balance} because $680$ is not a multiple of $12$.
Data points at diamond width of zero correspond to a standard spatial blocking scheme.

Our models are very accurate in predicting the code balance of corner-case stencil operators. There is a strong agreement between the model and the empirical results when the cache block (i.e., wavefront) fits in the L3 cache (in the range of 12--18\,\MiB). This shows that our implementation of the MWD blocking scheme can actually achieve the theoretical memory traffic reductions.
The measured code balance in Fig.~\ref{fig:code_balance} starts to deviate from the model at cache blocks larger than  about half the Intel Ivy Bridge's L3 cache size (i.e., 25\,\MiB). 
The deviation at this point can be predicted from our cache block size model,  considering the rule-of-thumb that half the cache size is usually usable for blocking.
This rule has emerged from experience for stencil codes that show an approximate balance of data volumes between the stencil array and other (streaming) data in the loop code. If the latter dominates the data volume, the factor of one
half may be reduced.

\subsection{Performance and energy consumption}

Thread scaling data of the performance and the measured energy
consumption for the three stencil operators under consideration are
shown in Figures~\ref{fig:7_pt_const_performance_energy},
\ref{fig:7_pt_var_performance_energy}, and
\ref{fig:25_pt_var_performance_energy}.
More detailed energy results, which separate CPU and DRAM power and energy,
are listed in Tables~\ref{table:7_pt_const_energy}, \ref{table:25_pt_var_energy}, and \ref{table:25_pt_var_energy} for those core counts which 
lead to lowest energy consumption.
Auto-tuning is used throughout the thread scaling experiments to find the performance-optimal set of parameters for each experiment.

We observe that tiling in $x$ does not achieve the desired performance improvements. The auto-tuner thus selects a full stride in $x$ in most of the experiments. We attribute this failure to the hardware prefetching unit
bringing in data beyond the block in $x$, which annihilates any advantage of the desired cache block size saving.

\subsubsection{7-point constant-coefficient stencil}

\begin{figure}[tbp]
    \centering
    \subfloat[Performance]{
        \centering
        \includegraphics[width=3.8cm]{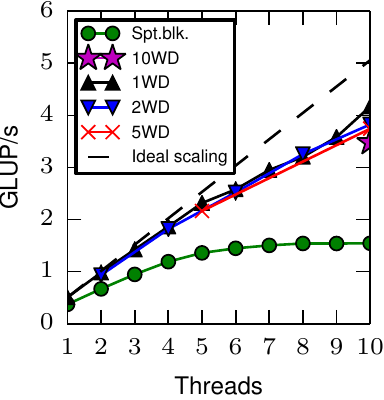}
        \label{fig:7_pt_const_performance}
	}
	\enskip
    \subfloat[Measured energy]{
        \centering
        \includegraphics[width=4.1cm]{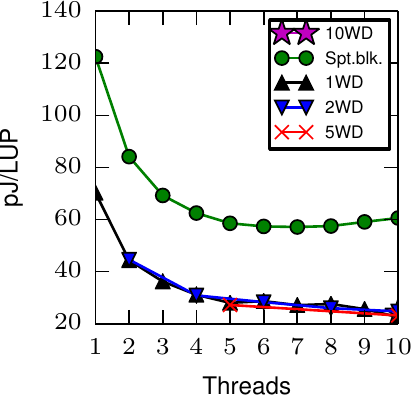}
        \label{fig:7_pt_const_energy}
	}
	\caption{7-point constant-coefficient stencil performance and energy. Grid size $N=960^3$.}
    \label{fig:7_pt_const_performance_energy}
\end{figure}

\begin{table}[tbp]
\centering
\begin{tabular}{|r|r||c|c|c|c|c|}
\hline
 & Method & Spt. Blk. & 1WD & 2WD & 5WD & 10WD \\
\cline{2-7}
           &    Threads &          6 &         10 &         10 &         10 &         10 \\
\cline{2-7}
           &     MLUP/s &       1448 &       4170 &       3825 &       3744 &       3481 \\
\hline
     Power &        CPU &      42.10 &      58.00 &      63.45 &      57.75 &      56.76 \\
\cline{2-7}
       [W] &       DRAM &      40.93 &      35.82 &      31.12 &      28.95 &      27.44 \\
\cline{2-7}
           &      Total &      83.03 &      93.81 &      94.56 &      86.70 &      84.20 \\
\hline
    Energy &        CPU &      29.09 &      13.92 &      16.59 &      15.42 &      16.31 \\
\cline{2-7}
  [pJ/LUP] &       DRAM &      28.28 &       8.60 &       8.14 &       7.73 &       7.88 \\
\cline{2-7}
           &      Total &      57.36 &      22.51 &      24.72 &      23.16 &      24.19 \\
\hline
\end{tabular}
\caption{7-point constant-coefficient stencil power dissipation and energy to solution}
\label{table:7_pt_const_energy}
\end{table}
This stencil operator shows the usual strongly saturating performance
with pure spatial blocking across the cores of the CPU chip (circles
in Fig.~\ref{fig:7_pt_const_performance}). We thus expect lowest
energy to solution at around six threads, which is confirmed by the
measurements (circles in Fig.~\ref{fig:7_pt_const_energy}).  The
corresponding column in Table~\ref{table:7_pt_const_energy} shows that
the overall power is almost evenly distributed between CPU and
DRAM. The variants with temporal blocking all have much better
performance, and thus CPU utilization, but exert less pressure on the
memory interface. Consequently, their CPU power is higher (between
58\,\W\ and 64\,\W) but their DRAM power is lower, with 10WD hitting
the minimum at just over 27\,\W. In fact, 10WD has the lowest overall
power of all WD codes.

Considering the power dissipation and performance numbers of all
variants it is evident that spatial blocking must have the largest
total energy consumption due to its low performance, and indeed the
last row of Table~\ref{table:7_pt_const_energy} proves that all WD
codes are more than a factor of two ahead in energy efficiency.  10WD,
although it leaves the CPU and DRAM ``cooler,'' is still considerably
slower than 1WD, which eventually causes a slightly worse energy
efficiency. Thus, the general rule of ``faster code is more energy
efficient'' seems to be confirmed for this case, but among the WD
versions this result emerges from a complex interplay of performance
and power dissipation between CPU and DRAM, and it may not hold for
other stencil operators.

Note that due to their reasonably good scalability 
(see Fig.~\ref{fig:7_pt_const_performance}), all WD codes
show minimum energy to solution at the full socket (10 cores).

\subsubsection{7-point variable-coefficient stencil}

\begin{figure}[tbp]
    \centering
    \subfloat[Performance]{
        \centering
        \includegraphics[width=3.9cm]{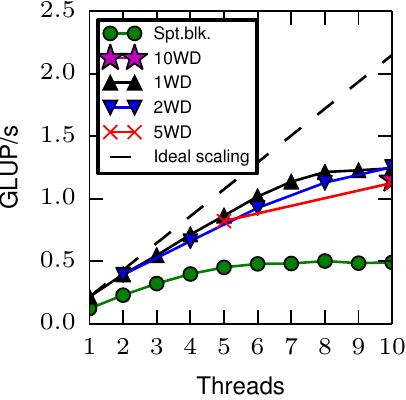}
        \label{fig:7_pt_var_performance}
	}
	\enskip
    \subfloat[Measured energy]{
        \centering
        \includegraphics[width=3.95cm]{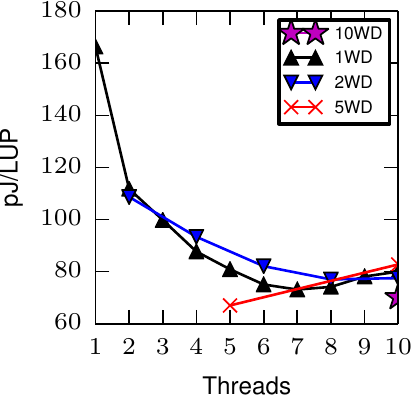}
        \label{fig:7_pt_var_energy}
	}
	\caption{7-point variable-coefficient stencil performance and energy. Grid size $N=680^3$.}
    \label{fig:7_pt_var_performance_energy}
\end{figure}

\begin{table}[tbp]
\centering
\begin{tabular}{|r|r||c|c|c|c|c|}
\hline
 & Method & Spt. Blk. & 1WD & 2WD & 5WD & 10WD \\
\cline{2-7}
           &    Threads &          6 &          8 &         10 &         10 &         10 \\
\cline{2-7}
           &     MLUP/s &        479 &       1214 &       1253 &       1126 &       1152 \\
\hline
     Power &        CPU &      39.78 &      48.26 &      59.19 &      54.11 &      52.93 \\
\cline{2-7}
       [W] &       DRAM &      47.40 &      41.66 &      37.94 &      38.73 &      26.91 \\
\cline{2-7}
           &      Total &      87.18 &      89.93 &      97.13 &      92.84 &      79.84 \\
\hline
    Energy &        CPU &      83.14 &      39.79 &      47.25 &      48.23 &      46.49 \\
\cline{2-7}
  [pJ/LUP] &       DRAM &      99.07 &      34.35 &      30.28 &      34.52 &      23.63 \\
\cline{2-7}
           &      Total &     182.21 &      74.14 &      77.53 &      82.76 &      70.12 \\
\hline
\end{tabular}
\caption{7-point variable-coefficient stencil power dissipation and energy to solution}
\label{table:7_pt_var_energy}
\end{table}
This stencil operator has the most unusual energy
vs.\ performance characteristic. First of all, due to the saturating
performance of 1WD the minimum energy operating point is at only eight
cores for this code (triangles in 
Fig.~\ref{fig:7_pt_var_performance_energy}). The 5WD variant, despite
showing about 30\% speedup from 5 to 10 cores, has lowest energy at
five cores (crosses). In practice one would usually not favor 
lower energy over shorter time to solution, which is why we have
included the 10-core power data in Table~\ref{table:7_pt_var_energy}.
5WD shows the energy advantage at 5 cores because it uses a diamond
width of $D_w=20$ compared to $D_w=8$ at 10 cores, resulting in
more savings in memory bandwidth, which leads to lower DRAM power. This
happens because more cache space is available per thread at 5 threads.
The performance advantage at 10 threads is not large enough
to compensate the larger DRAM and CPU power.

Comparing all variants with respect to energy consumption and
performance, we see that although 2WD has best performance, 
10WD shows lowest energy to solution, mainly due to its very 
low power dissipation in the DRAM. Since the
DRAM accounts for a significant fraction of the overall
power dissipation, the 7-point variable-coefficient stencil operator
with its high memory pressure (i.e., large code balance)
benefits not only in terms of performance but also in terms
of energy, especially when using 10WD which has the lowest
memory bandwidth utilization. However, its 6\% energy
advantage comes at the cost of an 8\% performance loss
compared to 2WD. 

Although tiling in $x$ is not generally improving the performance compared to~\cite{malas_mws_sisc2014}, it prevents the performance degradation of the 7-point variable-coefficient stencil at 10 threads in Fig.~\ref{fig:7_pt_var_performance}, where the best performance at 10 threads is achieved with a tile size of $340$ (half the leading dimension size). 

\subsubsection{25-point variable-coefficient stencil}

\begin{figure}[tbp]
    \centering
    \subfloat[Performance]{
        \centering
        \includegraphics[width=3.9cm]{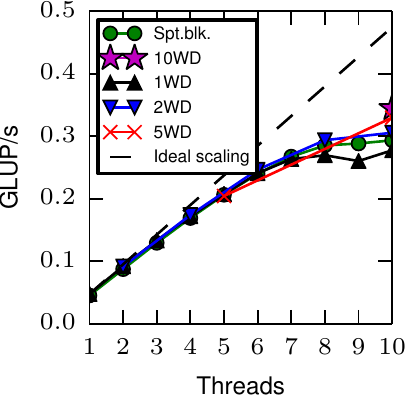}
        \label{fig:25_pt_var_performance}
	}
	\enskip
    \subfloat[Measured energy]{
        \centering
        \includegraphics[width=3.95cm]{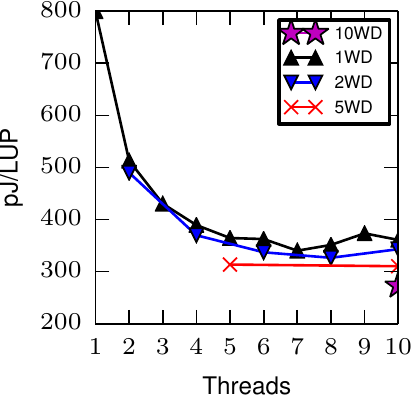}
        \label{fig:25_pt_var_energy}
	}
	\caption{25-point variable-coefficient stencil performance and energy. Grid size $N=480^3$.}
    \label{fig:25_pt_var_performance_energy}
\end{figure}

\begin{table}[tbp]
\centering
\begin{tabular}{|r|r||c|c|c|c|c|}
\hline
 & Method & Spt. Blk. & 1WD & 2WD & 5WD & 10WD \\
\cline{2-7}
           &    Threads &          8 &          7 &          8 &         10 &         10 \\
\cline{2-7}
           &     MLUP/s &        285 &        263 &        294 &        330 &        345 \\
\hline
     Power &        CPU &       46.1 &       44.1 &       51.2 &       53.8 &       53.3 \\
\cline{2-7}
       [W] &       DRAM &       48.5 &       45.5 &       44.7 &       48.4 &       40.7 \\
\cline{2-7}
           &      Total &       94.6 &       89.6 &       95.9 &      102.2 &       94.0 \\
\hline
    Energy &        CPU &      161.8 &      167.3 &      174.4 &      163.3 &      154.8 \\
\cline{2-7}
  [pJ/LUP] &       DRAM &      170.2 &      172.8 &      152.2 &      147.1 &      118.2 \\
\cline{2-7}
           &      Total &      331.9 &      340.2 &      326.5 &      310.4 &      273.0 \\
\hline
\end{tabular}
\caption{25-point variable-coefficient stencil power dissipation and energy to solution}
\label{table:25_pt_var_energy}
\end{table}
This is again a case where ``fastest'' also means ``least energy.''
The 10WD variant achieves best performance and lowest energy to
solution for this stencil operator (see
Fig.~\ref{fig:25_pt_var_performance_energy}), because it is the only
code that shows significant power savings in DRAM (see
Table~\ref{table:25_pt_var_energy}). 5WD, despite its 
substantial speedup from 5 to 10 cores, has the same energy 
to solution in both cases. The speedup is just barely sufficient
to compensate for the additional power from the larger number
of active cores.

\subsubsection{Code balance and energy consumption} 

Although these findings would not justify favoring 10WD over all other
options today and on the architecture under consideration, they show
clearly that the expected future trends towards more bandwidth-starved
systems and higher relative power dissipation in the memory subsystem
should be met with algorithms that exhibit lowest possible code balance.
This view is corroborated by another observation in our data: Across
all stencil operators, the \emph{overall} energy savings of temporal
blocking vs.\ standard spatial blocking are roughly accompanied by
equivalent runtime savings. But when the energy consumption of CPU and
DRAM are inspected separately it is evident that this equivalence 
emerges from the mutual cancellation of two opposing effects: 
While the CPU energy is less strongly correlated with the code 
performance, the DRAM energy shows an over-proportional reduction
for temporal blocking. 

\begin{figure}[tbp]
    \centering
    \subfloat[7-point constant-coefficient stencil at grid size $N=960^3$.]{
        \centering
        \includegraphics[width=6.0cm]{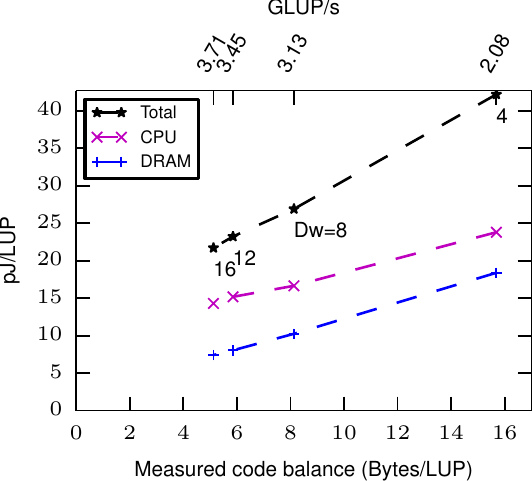}
        \label{fig:7_pt_const_code_balance_energy}
	}
	
    \subfloat[7-point variable-coefficient stencil at grid size $N=480^3$.]{
        \centering
        \includegraphics[width=6.2cm]{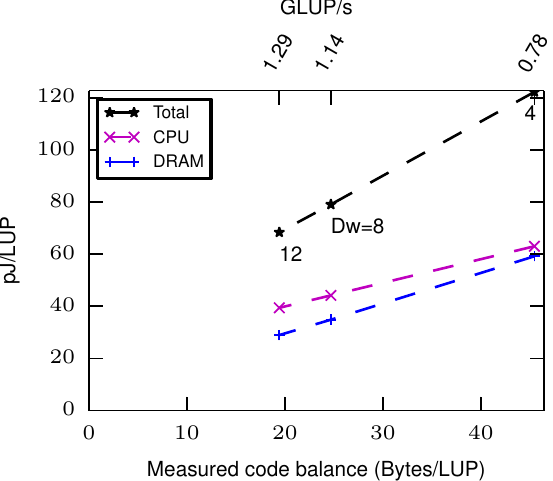}
        \label{fig:7_pt_var_code_balance_energy}
	}
	\caption{Energy vs.\ code balance for the seven-point stencils at 
          several diamond tile sizes, separately for DRAM and
          CPU and as a total sum. The corresponding performance of each experiment is shown on the top x-axis. The annotation at each point represents the used diamond width. 5WD is used in the experiments.}
    \label{fig:code_balance_energy}
\end{figure}

This can be seen more clearly in Fig.~\ref{fig:code_balance_energy}
where we have measured the energy to solution with respect to the code
balance for 5WD (as a consequence of setting different diamond tile
sizes) for both 7-point stencils (the diagram for the 25-point stencil
would only contain a single data point per set).  In both cases the
DRAM energy depends much more strongly on the code balance than the
CPU energy. This was expected from the observations described above,
but the CPU energy dependence is far from weak.  Overall there is an
almost linear dependence of energy on code balance, making the latter 
a good indicator of the former.

\section{Related work}

In 2009, Datta et al.~\cite{datta09} provide an exhaustive
review of the state of research on stencil code optimizations. They
cover the performance of several combinations of optimization
techniques, processors, and stencil operators. Of all optimization
options, temporal blocking is the most promising strategy since it
allows for a dramatic reduction of code balance and, potentially, a
decoupling from the memory bandwidth bottleneck. However, it is not
straightforward to apply since it requires careful handling of
inter-tile dependencies.  Many variants using different tile shapes
have been developed in recent years. Reviews can be found in Orozco
\textit{et al.}~\cite{orozco2011locality} and Zhou~\cite{Zhou2013}.
The idea of diamond tiling has been investigated by several groups,
e.g., Strzodka \textit{et al.}~\cite{Strzodka:2011:CAT},
Bandishti~\textit{et al.} \cite{Bandishti6468470}.

Wavefront-based blocking was introduced by Lamport~\cite{Lamport:1974}
and combined with tiling strategies by other authors
\cite{Strzodka:2011:CAT,Wonnacott845979,nguyen18993}. They all have in 
common that shared caches in modern multi-core processors are
not leveraged for improved cache reuse. This was first 
introduced by Wellein \textit{et al.}~\cite{wellein5254211}
with multicore wavefront temporal blocking. In recent work
\cite{malas_mws_sisc2014} we have combined this idea with 
previous work on diamond tiling to arrive at the MWD scheme
which is used in this work.

Our approach to studying performance behavior of optimized stencil
computations is based on a combination of auto-tuning (e.g., for
selecting appropriate diamond sizes) and model-guided performance
engineering, where we try to quantify the impact of bottlenecks on the
code execution much along the lines of the well-known Roof\/line
model~\cite{Williams09}. The code balance (inverse computational
intensity) is thus the ideal first-order metric for this.

The emergence of energy and energy-related metrics as new optimization
targets in HPC has sparked intense research on power issues in recent
years. For instance, the realization that low energy and low time to
solution may be opposing goals has led to activities in
multi-objective auto-tuning
\cite{bbalaprakash2013anl,Gschwandtner2014_euro_par}. However, 
there is very little work that tries to connect power dissipation
with code execution using simple synthetic models to gain 
insight without 
statistical or machine learning components. Hager et al.~\cite{CPE:CPE3180}
have constructed a simple power model that can explain
the main features of power scaling and energy to solution on
standard multicore processors. Choi et al.~\cite{Choi:2013:RME:2510661.2511392}
follow a slightly different approach by modeling the energy 
consumption of elementary operations such as floating-point 
operations and cache line transfers. In this work we try to
pick up some of those ideas to establish a connection between
energy consumption on the CPU and in the DRAM with the performance
and, more importantly, with the code balance of a stencil algorithm.

\section{Summary and outlook}

In this work we have established a memory traffic model for stencil
update schemes that are optimized by Multicore Wavefront Diamond (MWD)
temporal blocking. Our traffic model can predict the optimal code
balance as a function of the stencil radius, the diamond width, and
the number of domain-sized streams.  We have validated the model
predictions on a 10-core Intel Ivy Bridge by direct traffic
measurements for three stencil operators with different properties: a
7-point constant-coefficient stencil, a 7-point stencil with variable
coefficients, and a long-range 25-point stencil with variable
axis-symmetric coefficients. The model is very accurate if the
required cache block size (which is also predicted) fits into
about half the shared outer level cache size. This enables a useful
memory traffic calculation, and constitutes an important step towards 
improved model-guided automated tuning. 

By direct energy measurements for CPU and DRAM using the RAPL
facilities we could show that the DRAM power dissipation is a crucial
factor for energy to solution on the system under consideration, and
that it correlates strongly with the memory traffic. As a consequence,
the general observation that there is an almost linear dependence
between time and energy to solution may not always be true, even 
if the executed low-level code is identical.  Indeed we
have identified one case (7-point stencil operator with variable
coefficients) where the most time-efficient MWD variant is not the
most energy-efficient. Although these were not major deviations, the
observed dependence on the DRAM power points to expected future trend
towards a shift of power dissipation hot spots from execution resources
to data resources. Algorithms like MWD with strongly reduced memory
pressure stay abreast of these changes.

\begin{figure}[tbp]
    \centering
    \includegraphics[width=6.0cm]{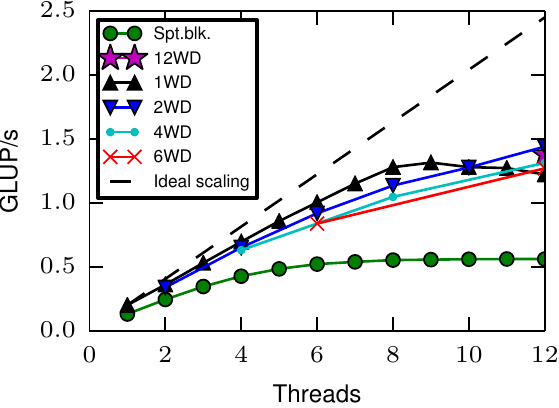}
    \caption{7-point variable-coefficient stencil thread scaling performance at a 12-core Intel Ivy bridge processor at Edison supercomputer using grid size $N=680^3$.}
    \label{fig:edison_7_pt_var_perf}
\end{figure}

We present some final performance
measurements in a different hardware setting
to substantiate this point:
Figure~\ref{fig:edison_7_pt_var_perf} shows the thread scaling
performance of the 7-point variable-coefficient stencil on a single
socket of a 12-core Ivy Bridge CPU in the  ``Edison''
system at NERSC. The socket has a measured STREAM TRIAD
bandwidth of $b_\mathrm S\approx 45\,\GBS$ and runs at
a base clock speed of 2.4\,\GHZ. 
Compared to the 10-core Intel Ivy Bridge we used in our earlier
experiments, this system has a 12.5\% higher memory bandwidth,
25\% more cores, and a 10\% higher clock speed.  As a result,
it is more bandwidth-starved, i.e., the ratio of memory bandwidth
to peak performance is lower. 
As a result, MWD (and especially 12WD) shows a significant
improvement in full-chip performance, much more than
spatial blocking (which can only benefit from the 12.5\%
bandwidth boost).

The recent availability of the new Intel Haswell-EP processor will add
a new twist to our study since the number of cores, peak performance,
cache sizes, and (to a lesser extent) the memory bandwidth will all be
increased considerably. We expect MWD to have even stronger advantages
in terms of performance and energy on this new architecture.

\section*{Acknowledgment}
The Extreme Computing Research Center at KAUST supported T. Malas.
Part of this work was supported by the German DFG priority programme
1648 (SPPEXA) within the project Terra-Neo,
and by the Bavarian Competence Network for Scientific High Performance
Computing in Bavaria (KONWIHR) under the project OMI4papps. We are indebted 
to Jan Treibig and Thomas R\"ohl (RRZE) who provided support 
with the LIKWID tool suite.
This research used resources of the National Energy Research
Scientific Computing Center, a DOE Office of Science User Facility 
supported by the Office of Science of the U.S. Department of Energy 
under Contract No. DE-AC02-05CH11231.

\bibliographystyle{IEEEtran}
\bibliography{references}

\end{document}